# Parity Anomalous Semimetal with Minimal Conductivity Induced by an In-Plane Magnetic Field

Binbin Wang[1†], Jiayuan Hu[1†], Bo Fu[2†], Jiaqi Li[1], Yunchuan Kong[1], Kai-Zhi Bai[3], Shun-Qing Shen[3*] and Di Xiao[1*]

[1]Huawei Technologies Co., Ltd., Shanghai, China

[2]School of Sciences, Great Bay University, Dongguan, China

[3]Department of Physics and State Key Laboratory of Optical Quantum Materials, The University of Hong Kong, Hong Kong, China

[†]These authors contribute equally to this work.

[*]Emails: xiaodi12@hisilicon.com; sshen@hku.hk

**Abstract**

The interplay between topological materials and local symmetry breaking gives rise to diverse topological quantum phenomena. A notable example is the parity-anomalous semimetal (PAS), which hosts a single unpaired gapless Dirac cone with a half-integer quantized Hall conductivity. Here, we realize this phase in a magnetic topological sandwich structure by applying an in-plane magnetic field. This configuration aligns the magnetization of one surface in-plane while preserving magnetization out-of-plane on the opposite surface, satisfying the condition for a gapless surface state near the Fermi level on only one surface. Our key evidence is a distinctive two-stage evolution of the conductivity tensor $(\sigma_{xy}, \sigma_{xx})$. The first stage culminates in the PAS at the fixed point $(\frac{e^2}{2h}, m\frac{e^2}{h})$, where $m \approx 0.6$ corresponds to the minimal longitudinal conductivity of a single gapless Dirac cone of fermions on a 2D lattice. This PAS state remains stabilized and is superposed with a gapped band flow in the second stage. This observation demonstrates that this state stabilized by the in-plane field resists localization—contrary to conventional expectations for 2D electron systems with broken time-reversal symmetry. The dynamic transition from an integer quantized insulator to a half-integer quantized semimetal establishes this material system as a versatile platform for exploring parity anomaly physics.

A high-mobility two-dimensional (2D) electron system with broken time-reversal symmetry leads to the integer quantum Hall effect (IQHE) when the Fermi level resides in the gap [1]. The crucial roles of both dissipationless chiral edge states and localized bulk states are what ultimately lead to the quantization of the Hall conductivity. Within this paradigm, a half-quantized Hall conductivity $\sigma_{xy} = \frac{e^2}{2h}$ signifies a critical point—an unstable fixed point in the renormalization group flow that separates two distinct insulating phases and is susceptible to perturbations like disorder. A profoundly distinct scenario emerges for Dirac fermions with certain broken symmetries. In odd space-time dimensions such as (2+1)-D, the parity symmetry breaking provides a condensed-matter pathway to circumvent the Nielsen-Ninomiya fermion doubling theorem [2,3], allowing for a single, unpaired gapless Dirac cone on a lattice. This situation is closely analogous to the (2+1)-dimensional parity anomaly in quantum field theory [4–6], and can result in a half-integer quantized Hall conductivity in the absence of a Landau level spectrum on a lattice [7–9]. Crucially, unlike the unstable critical point of an IQHE transition, the half-quantization in the parity-anomalous state defines a stable and metallic phase known as the parity anomalous semimetal (PAS). This stability is rooted in a topologically protected, gapless Dirac cone, which suppresses backscattering to confer inherent resilience against weak disorder and prevent localization [10–12].

A promising experimental platform for realizing the PAS is a semi-magnetic topological insulator (TI) bilayer structure, in which the Fermi level of one surface locates inside a magnetic gap while the other surface remains gapless, hosting an unpaired Dirac cone [8,9]. Experimental signature of half-quantized Hall conductivity $\sigma_{xy}$, accompanied by a finite longitudinal conductivity $\sigma_{xx}$, has recently been observed in doped or proximitized TIs [13,14]. However, such static configurations generally lead to a fixed value of $\sigma_{xy}$ with limited tunability, and the influence of $\sigma_{xx}$ remains unexplored, making the experimental verification of PAS somewhat ambiguous.

In this work, we fabricate magnetically doped TI sandwich structures with different dopants on the two surfaces which host two gapped Dirac cones of surface electrons, yielding distinct magnetic anisotropies with markedly different in-plane coercive fields. By applying an in-plane magnetic field to align the magnetization of one surface layer completely in-plane while the other surface's magnetization still largely out-of-plane, we achieve dynamic control over the phase transition from a Chern/Axion insulator to a PAS exhibiting a half-quantized $\sigma_{xy} = \frac{e^2}{2h}$ and a finite $\sigma_{xx}$, providing strong evidence for the parity anomaly state [15]. The distinct nature of this state is revealed in the two-stage evolution of the conductivity $(\sigma_{xy}, \sigma_{xx})$. In the first stage, one gap of the surface states closes, guiding the system to the PAS fixed point at $(\frac{e^2}{2h}, m\frac{e^2}{h})$, where $m \approx 0.6$. In the second stage, the trajectory reflects a superposition of this stabilized PAS and a tiny-gapped band flow, a signature markedly different from integer quantum Hall systems. Notably, the longitudinal conductivity $\sigma_{xx}$ consistently shows a minimal value near $m\frac{e^2}{h}$ when the Fermi level lies near the Dirac point of the single massless Dirac cone, a behavior reproduced across multiple samples. Temperature-driven evolution further reveals a trend toward this finite minimal value. Our results suggest the extended nature of the PAS, which, unlike localized conventional 2D electronic systems with broken time-reversal symmetry, ensures an intriguing minimal and possibly universal conductivity of the single massless Dirac cone at low temperatures. These findings offer new insight into the quantum criticality of the nondegenerate gapless Dirac states, and provide a context for understanding the strongly sample-dependent minimal conductivity observed in degenerate Dirac systems, such as monolayer graphene [16,17].

We fabricate a series of magnetically doped TI sandwich structures on SrTiO₃ (111) substrates using molecular beam epitaxy (MBE). As shown in Fig. 1(a), the heterostructure comprises three layers [18,19]: a 3-quintuple-layer (QL) $Cr_{0.19}(Bi,Sb)_{1.81}Te_3$ (CBST) bottom layer, a 10-QL undoped $(Bi,Sb)_2Te_3$ (BST) spacer, and a 3-QL $Cr_xV_{0.11}(Bi,Sb)_{1.89-x}Te_3$ (CVBST) top layer, where $x$ denotes the Cr concentration in the topmost CVBST layer. The chemical potential of the samples is tunable via the Bi:Sb ratio $\eta$, or by applying a bottom gate voltage. We set $\eta = 0.83$ so that the charge neutral point (CNP) is located near a gate voltage of $V_g = 0$. All data presented herein were measured at the CNP unless otherwise specified. Low-temperature magneto-transport measurements were performed in a dilution refrigerator with a base temperature of $T = 30$ mK, equipped with a 6-1-1 T vector magnet. Standard Hall bar devices, defined either by mechanical scratching or lithographic patterning, were aligned parallel to the $z = 0$ plane, with the current direction parallel to the $x$-axis, as illustrated in Fig. 1(b). Prior to each in-plane magnetic field ($B_\parallel$) sweep, the device was first initialized into a specific magnetization configuration (parallel or antiparallel) by applying an out-of-plane magnetic field ($B_\perp$). Details of sample preparations are provided in the Supplementary materials (SM) section i.

Following the initialization of Device A into a parallel magnetization configuration (Chern number $\mathcal{C} = \pm 1$), we sweep the in-plane field $B_\parallel$ up to $\pm 4$ T. As shown in Fig. 1(c), a shoulder-like feature emerges in the anomalous Hall conductivity (AHC) near $B_\parallel = \pm 1.0$ T, where $\sigma_{xy}$ plateaus at approximately $\frac{e^2}{2h}$. This feature signifies the PAS phase, which arises when the magnetization of the CBST layer is aligned in-plane, while that of the CVBST layer retains a sizable out-of-plane component. This specific magnetic configuration is schematically depicted in Fig. 1(c) as the PAS configuration [7,9]. Upon further increasing $B_\parallel$ to about $\pm 4$ T, the AHC drops to $\sigma_{xy} \approx 0$, indicating that both magnetizations are now fully in-plane. Minor deviations from ideal values are likely due to a slight misalignment between the device plane and the applied in-plane field, and/or imperfections in the Hall bar geometries.

By plotting the data of each sweep in the $(\sigma_{xy}, \sigma_{xx})$ conductivity plane, a distinct dip-like feature emerges near $(\sigma_{xy}, \sigma_{xx}) \approx (\pm \frac{e^2}{2h}, 0.6 \frac{e^2}{h})$, revealing a two-stage flow. This characteristic progression can be understood as follows: In the first stage, the increasing in-plane field closes the gap of the surface states on the bottom surface, driving a transition from a Chern insulator to a PAS. In the second stage, the gap on the top surface closes while the bottom surface remains gapless; the total conductivity flow is thus described by the sum of contributions from a massive Dirac band and the stable PAS. This flow transition is governed by the relationship between disorder and the band gaps. A gapped band contributes a non-zero $\sigma_{xx}$ and a non-quantized $\sigma_{xy}$ when disorder broadening exceeds its gap even at the CNP. The clear observation of the turning point at $(\pm \frac{e^2}{2h}, 0.6 \frac{e^2}{h})$ specifically requires that the disorder broadening remains much smaller than the gap of the still-gapped top surface band. This two-stage flow is a hallmark of our trilayer magnetic topological system and provides direct evidence for the PAS—a key signature that, to the best of our knowledge, has not been clearly demonstrated in previous renormalization group (RG) flow studies of magnetic topological insulators [13,20–22].

To probe the robustness of the critical PAS state against perturbations, we perform standard in-plane magnetic field ($B_\parallel$) sweeps while systematically varying several parameters, including a small out-of-plane field ($B_\perp$), the coercive field of the CVBST layer, the bottom gate voltage, and temperature. The robustness of PAS relies on an emergent parity (or vertical mirror) symmetry that protects its low-energy, gapless Dirac states. This symmetry is preserved under an in-plane magnetic field but is broken by an out-of-plane field. Given the high sensitivity of the CBST layer's magnetization direction to the out-of-plane field [23,24], a small $B_\perp$ can open a gap, driving a transition from the PAS into the Chern or axion insulating states. As shown in Fig. 2(a) for Device B, we start from a Chern insulator state ($\mathcal{C} = \pm 1$) and perform $B_\parallel$ sweeps under various $B_\perp$ fields ranging from 0 to $\pm 0.1$ T. For $|B_\perp| \leq 0.01$ T, the curves exhibit similar behavior, retaining the shoulder-like PAS feature near the half-quantized AHC value of $\sigma_{xy} \approx \frac{e^2}{2h}$. However, for $|B_\perp| > 0.01$ T, the outcome depends on the direction of $B_\perp$. If $B_\perp$ is opposite to the magnetization, the system first transitions into the axion insulating state ($\mathcal{C} = 0$), because increasing $B_\parallel$ destabilizes the CBST magnetic ordering, allowing the out-of-plane anisotropy to establish the antiparallel configuration [24]. Conversely, if $B_\perp$ is aligned with the magnetization, the PAS feature is suppressed, as the CBST magnetization can no longer be stabilized at moderate $B_\parallel$.

The RG flow diagrams on the $(\sigma_{xy}, \sigma_{xx})$ conductivity plane, shown in Fig. 2(b), clearly identify the PAS state near $(\pm \frac{e^2}{2h}, 0.6 \frac{e^2}{h})$. Although a perturbation with $|B_\perp| > 0.01$ T drives the system toward stable fixed points, the PAS state, identified by the characteristic dip-feature, remains robust as long as the PAS configuration is maintained. When $B_\parallel$ is further increased to 4 T, the system, now with gapless Dirac states on both surfaces, moves to a location near $(0, \frac{e^2}{h})$. Notably, the longitudinal conductivity at this point is not simply twice the value of $0.6 \frac{e^2}{h}$ observed for the PAS state. We also perform identical measurements on the same device but with the in-plane field applied perpendicular to the current direction, as shown in Fig. 2(c) and 2(d). The overall behavior is qualitatively similar to the parallel-field case, with the system consistently exhibiting the characteristic PAS shoulder/dip feature near $\sigma_{xy} \approx \pm \frac{e^2}{2h}$. However, at large $B_\parallel$, the flows converge near $(0, 1.2 \frac{e^2}{h})$, forming an upper semicircle centered at $(0, 0.6 \frac{e^2}{h})$. While the primary conductivity contributions from the two gapless Dirac states near the Dirac point at large $B_\parallel$ should, in principle, be identical, we attribute the observed difference of $\sim 0.2 \frac{e^2}{h}$ to the possible inter-surface coupling mediated by gapless side surfaces when $B_\parallel$ is parallel to the current, which is absent in the perpendicular case. Additional data measured on other samples show similar results, as shown in Fig. S5 and Fig. S6. This angle-dependent longitudinal conductivity variation has also been observed in CBC samples under a rotating in-plane magnetic field (Fig. S10). The origin of this discrepancy warrants further investigation. Nevertheless, the longitudinal minimal conductivity for the PAS state in both cases is the same at $\sim 0.6 \frac{e^2}{h}$.

To better understand the RG flows observed in Fig. 2, we carry out theoretical calculations to capture the essence of these behaviors (details in SM section xiv), as shown in Fig. 3. We compare the two representative scenarios where the out-of-plane field is zero or $0.1$ T. The masses of the top and bottom Dirac bands are labeled as $m_t$ and $m_b$, respectively, which can be modulated with an increasing in-plane magnetic field. For the first scenario, when $m_b$ becomes zero (blue curve in Fig. 3(a)) and $m_t$ (red curve) remains finite, there exists a region of in-plane field representing the stable PAS, as labeled on the $B_\parallel$ axis. For the second scenario, however, there is no such region for a stable PAS (Fig. 3(b)). The nearly perfect semi-circle between $(\frac{e^2}{h}, 0)$ and $(0,0)$ indicates the evolution of a single massive Dirac band from Chern insulator to axion insulator (or trivial insulator). This is because the $m_b$ curve quickly crosses zero and $m_t$ is relatively larger and has no contribution under the sweeping in-plane field and the small out-of-plane field. The theoretically calculated RG flow fits the experimental data quite well in both scenarios, which further confirms the robustness of the PAS state stabilized by its magnetization configuration.

The robustness of PAS also depends on the difference between the in-plane coercive fields of the two magnetic layers. As demonstrated in Fig. S7, the window for observing the PAS becomes narrower with increasing Cr concentration $x$ in the top layer. We grew a series of samples with $x$ varying from 0 to 0.19, which causes the coercive field of the CVBST layer to approach that of the CBST layer [25]. As expected, the shoulder-like PAS feature becomes less prominent with increasing $x$. Notably, however, the PAS feature from all different samples coincides near the same point in the conductivity plane with $\sigma_{xy} \approx \frac{e^2}{2h}$, $\sigma_{xx} \approx 0.6\frac{e^2}{h}$. This consistency suggests that for a PAS at the CNP, the Hall conductivity is robustly half-quantized, while the longitudinal conductivity may also assume a universal value.

The conductivity of the PAS also depends on the position of the Fermi level relative to the Dirac point. Our data suggest the minimal longitudinal conductivity occurs near CNP. We measure Device B by sweeping the bottom-gate voltage under different in-plane magnetic fields $B_{\parallel} = 0$ T (Chern insulator), 0.75 T (PAS), and 4 T (Dirac semimetal, both surfaces gapless), as shown in Fig. 4(a) and 4(b). The bottom gate voltage primarily tunes the Fermi level of the bottom CBST surface. For the Chern insulator, the Dirac point of the bottom surface is located near $V_g = 10$ V, characterized by $(\frac{e^2}{h}, 0)$. The position of the Fermi level stays at the Dirac point when $B_{\parallel}$ is varied. In the PAS state at $B_{\parallel} = 0.75$ T (determined from standard $B_{\parallel}$ sweeps), the AHC $\sigma_{xy}$ maintains a nearly half-quantized value of $\frac{e^2}{2h}$ across the entire gate range, as shown in Fig. 4(b), consistent with earlier reports in semi-magnetic systems [13]. The AHC reaches a maximum of about $0.58\frac{e^2}{h}$ near the Dirac point, which may be attributed to a minigap opening at the CVBST layer induced by interlayer magnetic coupling. Notably, the longitudinal conductivity $\sigma_{xx}$ exhibits a minimum of $\approx 0.65\frac{e^2}{h}$ at the Dirac point. This observation provides strong evidence for a minimal conductivity associated with a single unpaired Dirac cone. Although the value of $\sim 0.6\frac{e^2}{h}$ is consistent across all of our devices, whether or not it represents a universal value for the TI system requires future investigation. The PAS exists at the confluence of IQH criticality and Dirac fermion physics, where the predicted longitudinal conductivity is debated. For the IQH transition, theoretical values range from $\sigma_{xx} \approx 0.5 \frac{e^2}{h}$ in numerical studies [26] to $\frac{2e^2}{\pi h}$ from conformal field theory [27]. Separately, the minimal conductivity of massless Dirac fermions is non-universal and highly sensitive to disorder and theoretical approach (see SM Section xv). Our experiment bridges this divide, providing a concrete, measured value of $\sigma_{xx}$ that serves as a critical benchmark for future theoretical work. When both surfaces are gapless at $B_{\parallel} = 4$ T, $\sigma_{xy}$ drops to nearly zero, while $\sigma_{xx}$ shows a minimum value of $\approx 1.0 \frac{e^2}{h}$ at the Dirac point, resulting from the combined minimal conductivity of two connected gapless Dirac states from the top and bottom surfaces.

To investigate the finite minimal conductivity, we perform standard $B_{\parallel}$ sweeps on Device C at elevated temperatures up to 1.2 K, as shown in Fig. 4(c). The overall behavior of the flows remains similar, but the longitudinal conductivity increases with temperature, primarily due to the enhanced thermally activated transport between disordered magnetic domains. We extract the conductivities at the PAS state and plot them as a function of temperature in Fig. 4(d). The AHC $\sigma_{xy}$ remains nearly constant at $\sim \frac{e^2}{2h}$, demonstrating the robustness of the PAS phase as long as its magnetic configuration is maintained, even though the required $B_{\parallel}$ field becomes smaller at higher temperatures (See Fig. S8). In contrast, the longitudinal conductivity $\sigma_{xx}$ increases monotonically from approximately $0.67 \frac{e^2}{h}$ at base temperature to $0.81 \frac{e^2}{h}$ at 1.2 K. Additional data from other samples (Fig. S9) confirm the similar trend of a minimal $\sigma_{xx}$ at the lowest temperatures. This slow decrease of $\sigma_{xx}$ as temperature is lowered likely suggests a *finite* conductivity for this PAS state even in the zero-temperature limit. This observation is further supported by our theoretical calculations (fitting curves in Fig. 4(d)). Thus, our results suggest that the single, unpaired Dirac cone in the PAS could not be localized, even at absolute zero temperature. This behavior stands in stark contrast to that of conventional 2D electron systems with broken time-reversal symmetry, which invariably flow toward an insulating, localized fixed point [28,29].

In summary, our work provides compelling evidence for the realization of parity anomalous semimetal, characterized by a two-stage conductivity flow. The metallic nature of this state is confirmed by its temperature-independent longitudinal conductivity, which converges to a finite value of $\sigma_{xx} \approx 0.6 \, \frac{e^2}{h}$ at low temperatures. This PAS constitutes a novel experimental platform that intrinsically intertwines the universality of the integer quantum Hall transition with the electrodynamics of Dirac fermions. Our work bridges these two domains, establishing a crucial experimental benchmark to guide future research on topological quantum matter and refine theories of topological transport in critical systems.

*Note added.* —We have learned of an independent study by Zhuo *et al*. [30] on similar structures.

**Acknowledgements**

We acknowledge Huawei Semiconductor Failure Analysis Center Platform and Nanofab at Suzhou Institute of Nano-Tech and Nano-Bionics, CAS for the help of device fabrication and characterization. B.F. acknowledges the support from National Natural Science Foundation of China (Grants No.12504049), Guangdong Province Introduced Innovative R&D Team Program (Grant No. 2023QN10X136), Guangdong Basic and Applied Basic Research Foundation (Grant No. 2024A1515010430 and 2023A1515140008). S.-Q.S. acknowledges the support from Quantum Science Center of Guangdong-Hong Kong-Macao Greater Bay Area under Grant No. GDZX2301005 and the Research Grants Council, University Grants Committee, Hong Kong under Grants No. C7012-21G.

**Author contributions**

B.W., J.H. and B.F. contributed equally to this work. S.-Q.S and D.X. conceptualized the work; J.H. grew the samples; B.W., J.H. and J.L. fabricated the devices; B.W. conducted the transport measurements. B.F., K.-Z.B. and S.-Q.S. provided theoretical support. D.X., B.W., B.F., J.H. and S.-Q.S. analyzed the data and wrote the manuscript, with input from all authors.

**Competing interests**

The authors declare no competing interests.

**Data availability**

The data that support the plots within this study are available from the corresponding author upon reasonable request.

**Figures captions**

**Fig. 1 | Parity anomalous semimetal driven by an in-plane magnetic field in a magnetic topological insulator heterostructure. (a)** Schematic of the sandwich magnetic TI heterostructure with asymmetric magnetic doping. The $SrTiO_3$ substrate serves as the bottom gate dielectrics. **(b)** Illustration of the magnetotransport measurement setup for Hall bar devices. The in-plane field $B_\parallel$ is in parallel with the current direction (red arrow), while the out-of-plane field $B_\perp$ is perpendicular to the device plane. **(c)** Hall conductivity as a function of in-plane magnetic field for Device A. The device is initiate into $\pm 1$ Chern insulating state, then exhibit shoulder-like features near $B_\parallel = \pm 1.0$ T, where $\sigma_{xy}$ plateaus at approximately $\frac{e^2}{2h}$. These features correspond to the PAS magnetization configuration, as illustrated by the four schematics. **(d)** Renormalization group flow in $(\sigma_{xy}, \sigma_{xx})$ plane under sweeping in-plane field. This two-stage flow provides direct evidence for the PAS, as indicated by the red stars.

**Fig. 2 | Stability of the parity anomalous semimetal under small out-of-plane magnetic fields for Device B. (a)** Hall conductivity as a function of in-plane magnetic field in parallel with the current direction, under small out-of-plane magnetic fields. The shoulder-like feature persists for $|B_\perp| \leq 0.01$ T. **(b)** RG flow diagram of the sweeps in (a). Both semicircles at the bottom half of the diagram identify the PAS state near $(\pm \frac{e^2}{2h}, 0.6\frac{e^2}{h})$. **(c)** Hall conductivity as a function of in-plane magnetic field perpendicular to the current direction, under small out-of-plane magnetic fields, exhibiting similar behaviors as in (a). **(d)** RG flow diagram of the sweeps in (c), similarly identifying the PAS state near $(\pm \frac{e^2}{2h}, 0.6\frac{e^2}{h})$, while having a slightly larger longitudinal conductivity ($\sim 1.2\frac{e^2}{h}$) at large in-plane magnetic fields.

**Fig. 3 | Theoretical analysis of the flows with or without the out-of-plane magnetic**

**field.** **(a)** Top panel: masses of the top ($m_\text{t}$, red curve) and bottom ($m_\text{b}$, blue curve) surface layers as functions of in-plane magnetic field without out-of-plane perturbation. There is a significant region on the $B_\parallel$ axis identified as PAS. CI and DS stand for Chern insulator and Dirac semimetal, respectively. Bottom panel: The data and theoretical fitting of the RG flow of the in-plane sweep with $B_\perp = 0$. **(b)** Top panel: masses of the top ($m_\text{t}$, red curve) and bottom ($m_\text{b}$, red curve) surface layers as functions of in-plane magnetic field with $B_\perp = 0.1$ T, which drives the system into the axion insulating (AI) state. Bottom panel: The data and theoretical fitting of the RG flow of the in-plane sweep with $B_\perp = 0.1$ T.

**Fig. 4 | Minimal longitudinal conductivity at the CNP, and its temperature dependence** **(a)** The bottom gate dependence of the longitudinal conductivity of the Chern insulator (blue, $B_\parallel = 0$), the PAS (red, $B_\parallel = 0.75$ T), and both surfaces gapless (green, $B_\parallel = 4.0$ T). They all show a minimum at the Dirac point. **(b)** The bottom gate dependence of the Hall conductivity of the Chern insulator (blue, $B_\parallel = 0$), the PAS (red, $B_\parallel = 0.75$ T), and both surfaces gapless (green, $B_\parallel = 4.0$ T). The measurements in (a) and (b) were taken on Device B. **(c)** The renormalization group flow diagram of the in-plane field sweeps at different temperatures for Device C. **(d)** The extracted temperature dependent longitudinal and Hall conductivities of the PAS state, which exhibit an almost constant $\sigma_{xy} = \frac{e^2}{2h}$ and a slowly decreasing trend of $\sigma_{xx}$. The dashed line fittings are obtained through theoretical calculations, see Supplementary materials section xiii.

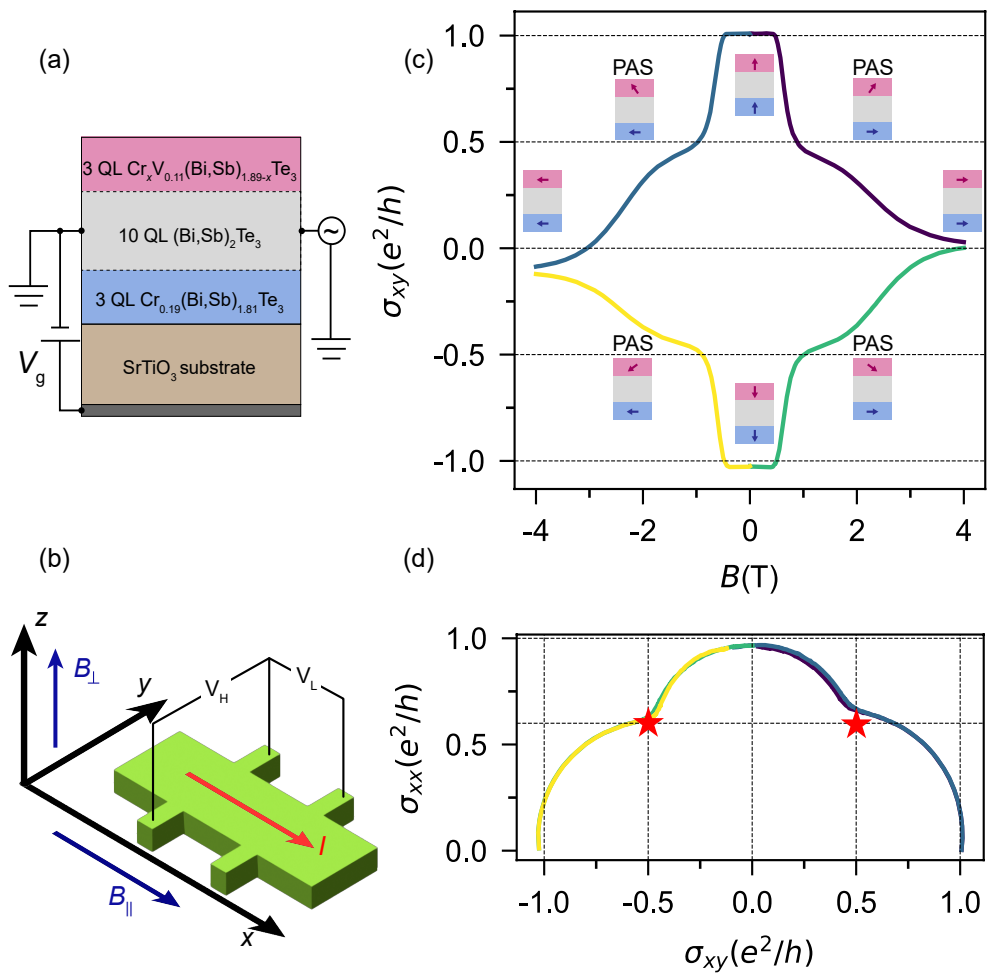

(a) 3 QL Cr$_x$V$_{0.11}$(Bi,Sb)$_{1.89-x}$Te$_3$
10 QL (Bi,Sb)$_2$Te$_3$
3 QL Cr$_{0.19}$(Bi,Sb)$_{1.81}$Te$_3$
SrTiO$_3$ substrate

$V_g$

(b)
$z$
$B_\perp$
$y$    $V_H$    $V_L$
$I$
$B_\parallel$
$x$

(c) $\sigma_{xy}(e^2/h)$
PAS    PAS
PAS    PAS
$B$(T)

(d) $\sigma_{xx}(e^2/h)$
$\sigma_{xy}(e^2/h)$

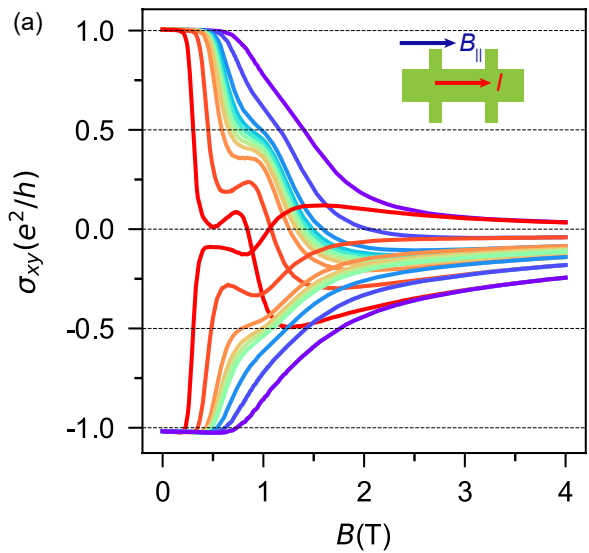

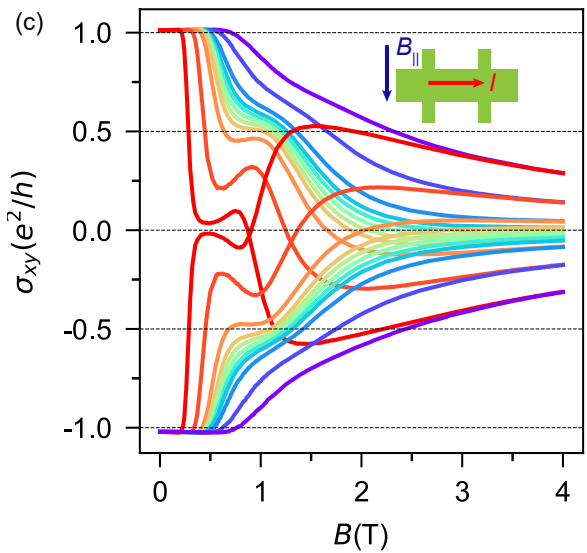

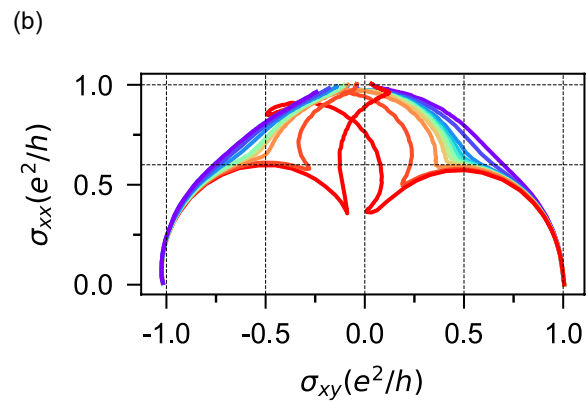

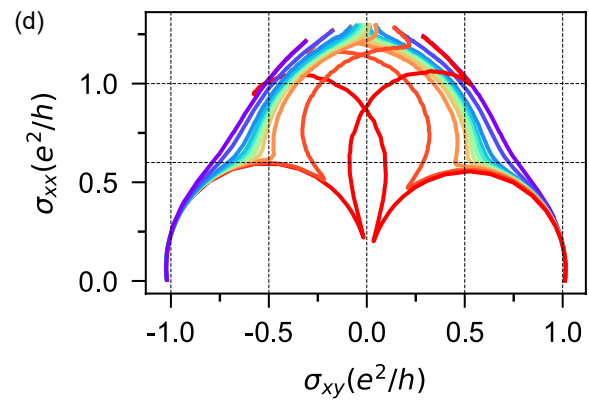

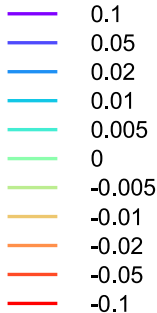

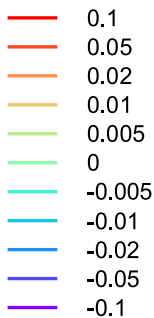

From $C=1$, $B_\perp(T)=$
- 0.1
- 0.05
- 0.02
- 0.01
- 0.005
- 0
- −0.005
- −0.01
- −0.02
- −0.05
- −0.1

From $C=-1$, $B_\perp(T)=$
- 0.1
- 0.05
- 0.02
- 0.01
- 0.005
- 0
- −0.005
- −0.01
- −0.02
- −0.05
- −0.1

**(a)**

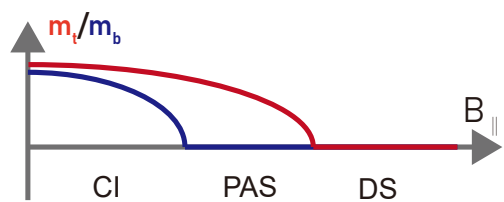

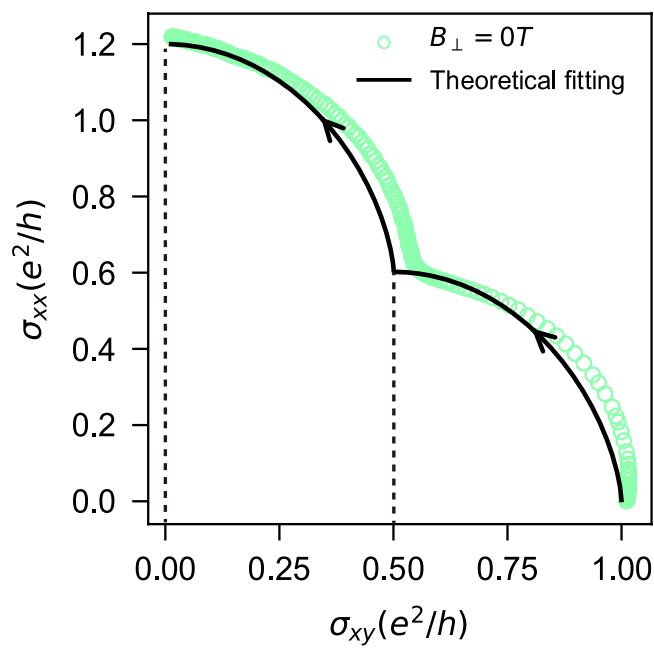

**(b)**

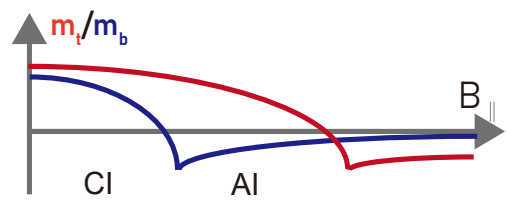

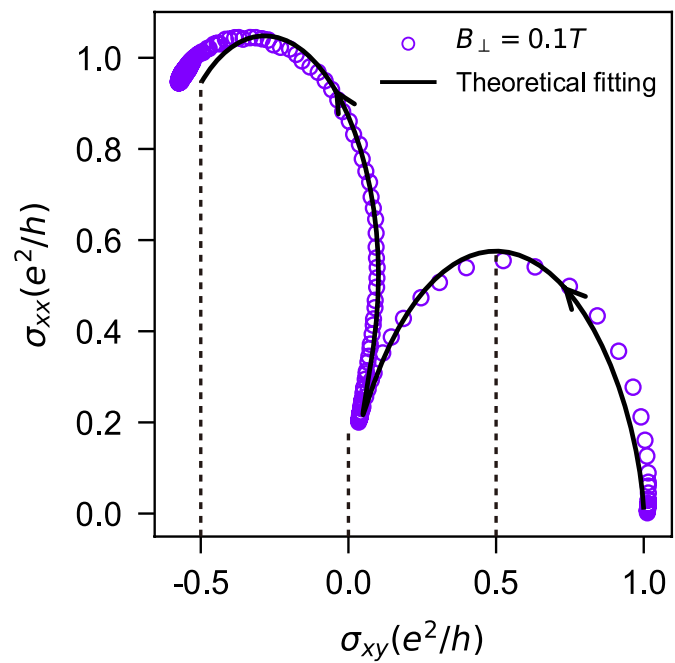

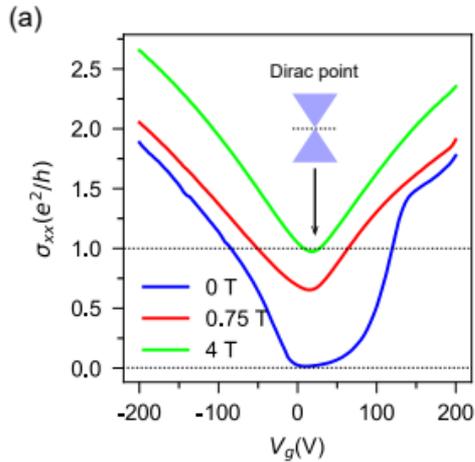

(a)

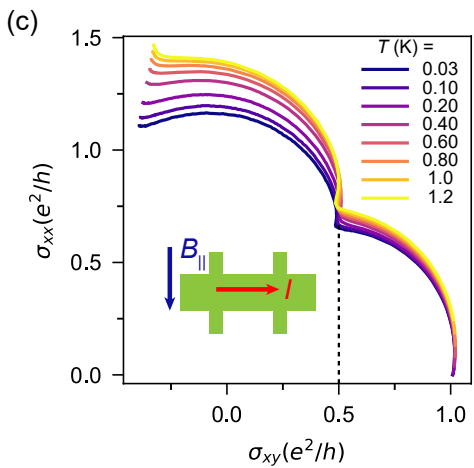

(c)

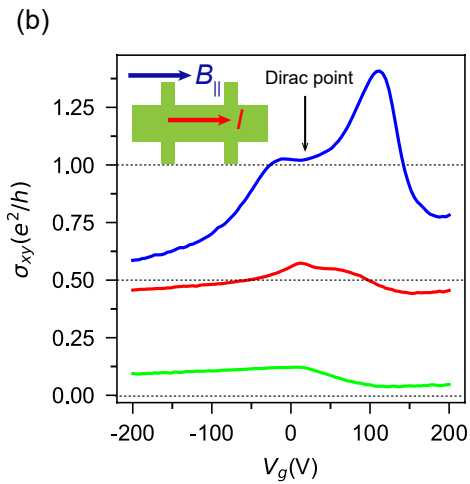

(b)

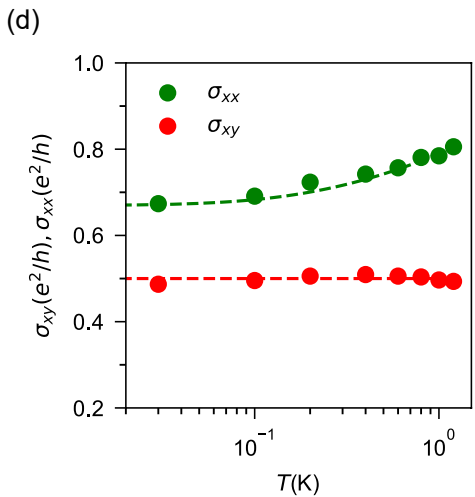

(d)